\documentclass[graybox,vecphys]{svmult}

\usepackage{type1cm}        
\usepackage{makeidx}         
\usepackage{graphicx}        
\usepackage{multicol}        
\usepackage[bottom]{footmisc}

\usepackage{newtxtext}       %
\usepackage{newtxmath}       
\usepackage{subfigure}

\usepackage{bm}

\makeindex             

\newcommand{\tauw}{\tau_{\mathrm w}}
\newcommand{\oltauw}{\overline{\tau}_{\mathrm w}}
\newcommand{\kn}{\mathrm{Kn}}

\begin{document}

\title*{A generalized slip-flow theory for a slightly rarefied gas flow induced by discontinuous wall temperature}
\titlerunning{A generalized slip-flow theory for a slightly rarefied gas flow}
\author{Satoshi Taguchi and Tetsuro Tsuji}
\institute{Satoshi Taguchi \at Department of Advanced Mathematical Sciences, Graduate School of Informatics, Kyoto University, Kyoto 606-8501, Japan \email{taguchi.satoshi.5a@kyoto-u.ac.jp}
\and Tetsuro Tsuji \at Department of Advanced Mathematical Sciences, Graduate School of Informatics, Kyoto University, Kyoto 606-8501, Japan \email{tsuji.tetsuro.7x@kyoto-u.ac.jp}}
%
%
\maketitle

\abstract*{A system of fluid-dynamic-type equations and their boundary conditions derived from a system of the Boltzmann equation is of great importance in kinetic theory when we are concerned with the motion of a slightly rarefied gas. It offers an efficient alternative to solving the Boltzmann equation directly and, more importantly, provides a clear picture of the flow structure in the near-continuum regime. However, the applicability of the existing slip-flow theory is limited to the case where both the boundary shape and the kinetic boundary condition are smooth functions of the boundary coordinates, which precludes, for example, the case where the kinetic boundary condition has a jump discontinuity. In this paper, we discuss the motion of a slightly rarefied gas caused by a discontinuous wall temperature in a simple two-surface problem and illustrate how the existing theory can be extended. The discussion is based on our recent paper [Taguchi and Tsuji, J. Fluid Mech. 897, A16 (2020)] supported by some preliminary numerical results for the newly introduced kinetic boundary layer (the Knudsen zone), from which a source-sink condition for the flow velocity is derived.}

\abstract{A system of fluid-dynamic-type equations and their boundary conditions derived from a system of the Boltzmann equation is of great importance in kinetic theory when we are concerned with the motion of a slightly rarefied gas. It offers an efficient alternative to solving the Boltzmann equation directly and, more importantly, provides a clear picture of the flow structure in the near-continuum regime. However, the applicability of the existing slip-flow theory is limited to the case where both the boundary shape and the kinetic boundary condition are smooth functions of the boundary coordinates, which precludes, for example, the case where the kinetic boundary condition has a jump discontinuity. In this paper, we discuss the motion of a slightly rarefied gas caused by a discontinuous wall temperature in a simple two-surface problem and illustrate how the existing theory can be extended. The discussion is based on our recent paper [Taguchi and Tsuji, J. Fluid Mech. 897, A16 (2020)] supported by some preliminary numerical results for the newly introduced kinetic boundary layer (the Knudsen zone), from which a source-sink condition for the flow velocity is derived.}


\section{\label{sec:intro}Introduction}
Let us consider a rarefied gas in contact with a smooth boundary (or boundaries). We are concerned with the steady behavior of the gas. Suppose that the molecular mean free path is small compared with the characteristic system size (the Knudsen number is small). Then, it is often advantageous to solve the fluid-dynamic system derived from the Boltzmann system. This approach is known as the generalized slip-flow theory and was developed notably by Sone and his coworkers \cite{Sone69,Sone71,Sone02,Sone07}. 

The generalized slip-flow theory is based on the asymptotic analysis of the Boltzmann system for small Knudsen numbers. Both the boundary shape and the boundary condition need to be smooth. This smoothness condition is required for the Knudsen-layer problem to be reduced to a half-space problem of a kinetic equation in space one dimension, from which the slip/jump boundary conditions are obtained. 

The smoothness condition can be, however, restrictive in some situations. For example, S.T. considered in \cite{Taguchi-Aoki_JFM12} a rarefied gas flow around a sharp edge with different surface temperatures on each side. But due to the limitation, only a qualitative argument was possible for the flow structure around the edge. Motivated by this, in this article, we discuss the possibility to extend the generalized slip-flow theory to the case where the boundary condition has a jump discontinuity in a simple two-surface problem. That is, we consider a steady rarefied gas flow between two parallel plates with a discontinuous wall temperature in the framework of the generalized slip-flow theory. The discussion is based on our recent paper \cite{taguchi_tsuji_JFM_2020} with some new numerical result, which supports the present theory.

Finally, we remark on the following. In our problem (to be stated next), the boundary condition has a jump discontinuity (through the plate's temperature distribution). This induces discontinuities of the velocity distribution function on the boundary, and they propagate into the gas region. This feature is important in a numerical analysis and was taken into account in \cite{Aoki-Takata-Aikawa-Golse_PHF97}, where a similar temperature-driven flow has been considered (see also \cite{taguchi_tsuji_JFM_2020}). It is also considered in our numerical results shown in Sect.~\ref{sec:numerical}, although the numerical approach is different. The propagation of boundary-induced discontinuity in kinetic equations is also a mathematical concern and has been investigated in, e.g., \cite{Aoki-Bardos-Dogbe-Golse_01,Kim_2011,esposito-guo-kim-marra_2013,Guo-Kim-Tonon-Trescases_2016,Kawagoe-Chen_JSP2018}. 

\section{\label{sec:problem_and_formulation}Problem and formulation}
\subsection{\label{subsec:problem}Problem}
\begin{figure}
    \centering
    \includegraphics[width=0.5\textwidth]{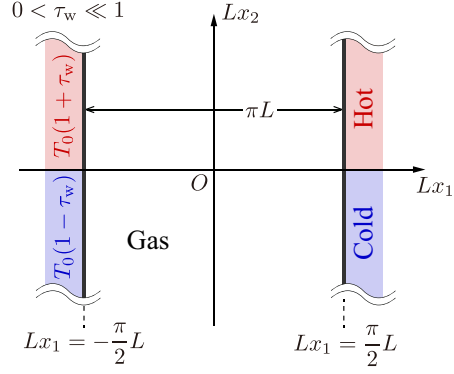}
    \caption{Schematic of the problem. A rarefied gas between two parallel plates located at $x_1=\pm \pi/2$ with a step-like temperature distribution is considered. The temperature of the plates is discontinuous at $x_2=0$.}
    \label{fig:problem}
\end{figure}

Let $L$ be the reference length and let $\rho_0$, $T_0$, and $p_0$ be the reference density, temperature, and pressure of the gas, respectively. We consider a monatomic rarefied gas occupying the space between two parallel plates located at $x_1 = -\frac{\pi}{2}$ and $x_1=\frac{\pi}{2}$, where $(L x_1,L x_2,L x_3)$ is the Cartesian coordinates, as shown in Fig.~\ref{fig:problem}. The upper halves of the plates ($x_2 > 0$) are kept at temperature $T_0(1 + \tauw)$, while the lower halves ($x_2 < 0$) at temperature $T_0(1 - \tauw)$, where $\tauw$ is a constant. Henceforth, we assume $\tauw>0$. Therefore, the surfaces' temperature has a step-like distribution, which is discontinuous at $x_2 = 0$ with the jump $2T_0 \tauw$. We also assume that the gas is subject to no pressure gradient nor external force. We investigate the steady behavior of the gas under the following assumptions: (i) the behavior of the gas is described by the Boltzmann equation; (ii) the gas molecules make diffuse reflection on the plates; (iii) $\tauw$ is so small that the equation and boundary conditions can be linearized around the reference equilibrium state at rest with density $\rho_0$ and temperature $T_0$; (iv) 
the Knudsen number defined by the molecular mean free path at the reference state divided by $L$ is small.


\subsection{Formulation}
Let us denote by $(2RT_0)^{1/2} (\zeta_1,\zeta_2,\zeta_3)$ the molecular velocity ($R$ is the specific gas constant) and by $\rho_0 (2RT_0)^{-3/2} (1+ \phi(\vec{x},\vec{\zeta}))E$ the velocity distribution function, where $E=\pi^{-3/2} \exp(-|\vec{\zeta}|^2)$. The time-independent Boltzmann equation reads
\begin{align}
&
\zeta_i \partial_i \phi = \frac{1}{\varepsilon} \mathscr{L}(\phi),
\label{e:LB_2}
\end{align}
where $\partial_i = \partial/\partial x_i$, $\mathscr{L}$ is the linearized collision operator \cite{Sone07},
and $\varepsilon$ is a parameter 
defined by
\begin{equation*}
    \varepsilon = \frac{\sqrt{\pi}}{2} \kn = \frac{\sqrt{\pi}}{2} \frac{\ell_0}{L}
    \qquad (\text{Kn: Knudsen number}).
\end{equation*}
Here, $\ell_0$ is the mean free path of the gas molecules in the equilibrium state at rest with temperature $T_0$ and density $\rho_0$. Note that $\varepsilon$ is the Knudsen number multiplied by $\sqrt{\pi}/2$.
The operator $\mathscr{L}$ is given by
\begin{subequations}
\begin{align}
& \mathscr{L}(F) = \int_{(\vec{\zeta}_*,\vec{e}) \in \mathbf{R}^3 \times \mathbf{S}^2} E_* (F'_* + F' - F_* - F) \,B \,\D\Omega(\vec{e}) \D\vec{\zeta}_*,
\\
& F = F(\vec{\zeta}), \quad F_* = F(\vec{\zeta}_*), \quad F' = F(\vec{\zeta}'), \quad F_*' = F(\vec{\zeta}_*'),  
\\
& \vec{\zeta}' = \vec{\zeta} + [(\vec{\zeta}_* - \vec{\zeta}) \cdot \vec{e}] \vec{e},
\quad \vec{\zeta}_*' = \vec{\zeta}_* - [(\vec{\zeta}_* - \vec{\zeta}) \cdot \vec{e}] \vec{e},
\\
& B=B\left(\frac{|\vec{e}\cdot(\vec{\zeta}_* - \vec{\zeta})|}{|\vec{\zeta}_* - \vec{\zeta}|},|\vec{\zeta}_*-\vec{\zeta}|\right),
\quad E_* = \frac{1}{\pi^{3/2}}e^{-|\vec{\zeta}_*|^2},
\end{align}
\end{subequations}
where $\D\Omega(\vec{e})$ is the solid angle element in the direction of $\vec{e}$, 
$B$ is a non-negative function whose functional form is determined by the designated intermolecular force. For example, $B = \frac{1}{4\sqrt{2\pi}}|\vec{e}\cdot (\vec{\zeta}_* - \vec{\zeta})|$ for a hard-sphere gas.
The diffuse reflection boundary conditions on the plates are summarized as
\begin{subequations}\label{e:diffuse_2}
\begin{align}
&
\phi = 2 \sqrt{\pi} \int_{\zeta_1 < 0} |\zeta_1| \phi E \D \vec{\zeta}
 \pm (|\vec{\zeta}|^2 - 2) \tau_{\mathrm{w}}, \quad \zeta_1 > 0
\qquad \left(x_1 = - \frac{\pi}{2}, \ x_2 \gtrless 0\right),
\label{e:diffuse_2a}
\\
&
\phi = 2 \sqrt{\pi} \int_{\zeta_1 > 0} |\zeta_1| \phi E \D \vec{\zeta}
 \pm (|\vec{\zeta}|^2 - 2) \tau_{\mathrm{w}}, \quad \zeta_1 < 0
\qquad \left(x_1 = \frac{\pi}{2}, \ x_2 \gtrless 0\right),
\label{e:diffuse_2b}
\end{align}
\end{subequations}
where $\D\vec{\zeta} = \D\zeta_1 \D\zeta_2 \D\zeta_3$. 

The macroscopic quantities of interest, namely, the density, the flow velocity, the temperature, and the pressure of the gas denoted by $\rho_0 (1 + \omega)$, $(2RT_0)^{1/2} u_i$, $T_0(1 + \tau)$, and $p_0 ( 1 + P)$, respectively, are defined in terms of $\phi$ as
\begin{align}
\omega &= \langle \phi \rangle,
\quad
u_i = \langle \zeta_i \phi \rangle,
\quad
\tau = \frac{2}{3} \left\langle \left(|\vec{\zeta}|^2 - \frac{3}{2}\right) \phi \right\rangle,
\quad
P = \frac{2}{3} \langle |\vec{\zeta}|^2 \phi \rangle = \omega + \tau,
\end{align}
where $\langle \cdot \rangle$ designates 
\begin{equation}
\langle F \rangle = \int_{\mathbf{R}^3} F(\vec{\zeta}) E \D\vec{\zeta}.
\end{equation}

In the present two-dimensional problem, we may assume that $\phi$ is independent of $x_3$.
Nevertheless, the $x_3$-dependency has not been precluded in the above formulation for later convenience.

The study on the behavior of a slightly rarefied gas (i.e., the gas with small Knudsen numbers) has a long history (see, e.g., references in \cite{Sone02}). In the case of a smooth boundary, Sone and his coworkers have extensively studied the question both for the steady \cite{Sone69,Sone71,Sone02, Sone07} and unsteady \cite{Sone07,Takata-Hattori-JSP-2012} settings. It is based on the asymptotic analysis of the Boltzmann system for small Knudsen numbers, and the theory is nowadays known as the generalized slip-flow theory. However, the approach above precludes the discontinuous boundary data. 
One of the paper's purposes is to show that we can extend Sone's asymptotic theory to include the latter situation.

\section{\label{sec:smooth}Case of a smooth temperature distribution}
Before we discuss the discontinuous surface temperature case, it is useful to review the case of a smooth temperature distribution. Let the temperature of the two plates be given by $T_0(1 + \oltauw)$, where $\oltauw$ is a smooth function of $(x_2,x_3)$. Then, assuming the diffuse reflection condition, the boundary conditions \eqref{e:diffuse_2a} and \eqref{e:diffuse_2b} are replaced by
\begin{align}
\phi & = 2 \sqrt{\pi} \int_{\zeta_1 \lessgtr 0} |\zeta_1| \phi E \D\vec{\zeta} + (|\vec{\zeta}|^2 - 2) \oltauw,
\quad \zeta_1 \gtrless 0
\nonumber \\
& \qquad \left(x_1 = \mp \frac{\pi}{2},\ -\infty < x_2 < \infty, \ -\infty < x_3 < \infty \right).
\label{e:diffuse_smooth}
\end{align}
We consider the asymptotic behavior of the solution $\phi$ of the linear system \eqref{e:LB_2} and \eqref{e:diffuse_smooth} for small $\varepsilon$ following Sone's method \cite{Sone02,Sone07}. It should be noted that for the linearization, $|\partial_i \oltauw| \ll 1$ should be assumed.

By the symmetry of the problem, one can assume that the solution is even with respect to $x_1=0$. Therefore, in the sequel, we consider the problem only in the left-half domain $D^-=\{(x_1,x_2,x_3)\, | -\frac{\pi}{2} < x_1 < 0, \, -\infty<x_2<\infty,\,-\infty<x_3<\infty \}$. The solution in the right-half domain is obtained from that of $D^-$ by $\phi(x_1,x_2,x_3,\zeta_1,\zeta_2,\zeta_3)=\phi(-x_1,x_2,x_3,-\zeta_1,\zeta_2,\zeta_3)$.

According to \cite{Sone02}, the solution is expressed in the form
\begin{align}
\phi = \phi_{\mathrm H} + \phi_{\mathrm K},
\end{align}
where $\phi_{\mathrm H}$ is called the Hilbert solution and describes the overall behavior of the gas,
while $\phi_{\mathrm K}$ is a correction to $\phi_{\mathrm H}$ required in the vicinity of the boundary (the Knudsen-layer correction). More precisely, $\phi_{\mathrm H}$ is a solution to Eq.~\eqref{e:LB_2} subject to the condition $\partial_i \phi_{\mathrm H} = O(\phi_{\mathrm{H}})$ (i.e., moderately varying solution). On the other hand, $\phi_{\mathrm{K}}$ is appreciable only in a thin layer (the Knudsen layer) adjacent to the boundary $x_1=-\frac{\pi}{2}$, whose thickness is of the order of $\varepsilon$. The Knudsen-layer correction $\phi_{\mathrm{K}}$ is subject to the conditions
\begin{align}
\partial_1 \phi_{\mathrm{K}} = O(\phi_{\mathrm{K}}/\varepsilon),
\quad (\delta_{ij} - n_i n_j) \partial_j \phi_{\mathrm{K}} = O(\phi_{\mathrm{K}}),
\end{align}
where $\delta_{ij}$ is Kronecker's delta and $\vec{n}=(1,0,0)$.
The $\phi_{\mathrm{H}}$ and $\phi_{\mathrm{K}}$ are expanded in $\varepsilon$ as
\begin{subequations}
\begin{align}
\phi_{\mathrm{H}} & = \phi_{\mathrm{H}0} + \varepsilon \phi_{\mathrm{H}1} + \varepsilon^2 \phi_{\mathrm{H}2} + \cdots,
\\
\phi_{\mathrm{K}} & = \varepsilon \phi_{\mathrm{K}1} + \varepsilon^2 \phi_{\mathrm{K}2} + \cdots.
\end{align}
\end{subequations}
Accordingly, the macroscopic quantities $h$ ($h=\omega,\,u_i,\,\tau,\,P$) are also expressed as
\begin{subequations}
\begin{align}
& h = h_{\mathrm{H}} + h_{\mathrm{K}},
\\
& h_{\mathrm{H}} = h_{\mathrm{H}0} + \varepsilon h_{\mathrm{H}1} + \varepsilon^2 h_{\mathrm{H}2} + \cdots,
\\
& h_{\mathrm{K}} = \varepsilon h_{\mathrm{K}1} + \varepsilon^2 h_{\mathrm{K}2} + \cdots,
\end{align}
\end{subequations}
where
\begin{subequations}
\begin{align}
& \omega_{\mathrm{H}m} = \langle \phi_{\mathrm{H}m} \rangle,
\quad
u_{i\mathrm{H}m} = \langle \zeta_i \phi_{\mathrm{H}m} \rangle,
\quad
\tau_{\mathrm{H}m} = \frac{2}{3} \left\langle \left(|\vec{\zeta}|^2 - \frac{3}{2}\right) \phi_{\mathrm{H}m} \right\rangle,
\\
&  P_{\mathrm{H}m} = \omega_{\mathrm{H}m} + \tau_{\mathrm{H}m},
\end{align}
\end{subequations}
($m = 0,1,\ldots$), and 
\begin{subequations}
\begin{align}
& \omega_{\mathrm{K}m} = \langle \phi_{\mathrm{K}m} \rangle,
\quad
u_{i\mathrm{K}m} = \langle \zeta_i \phi_{\mathrm{K}m} \rangle,
\quad
\tau_{\mathrm{K}m} = \frac{2}{3} \left\langle \left(|\vec{\zeta}|^2 - \frac{3}{2}\right) \phi_{\mathrm{K}m} \right\rangle,
\\
& P_{\mathrm{K}m} = \omega_{\mathrm{K}m} + \tau_{\mathrm{K}m},
\end{align}
\end{subequations}
($m=1,2,\cdots$).

Then, it is shown in \cite{Sone02} that $\phi_{\mathrm{H0}}$, $\phi_{\mathrm{H}1}$, and $\phi_{\mathrm{K}1}$ are expressed in the form
\begin{subequations}
\begin{align}
& \phi_{\mathrm{H}0} = \phi_{\mathrm{eH}0}, \label{e:phi_H0} \\
& \phi_{\mathrm{H}1} = \phi_{\mathrm{eH}1} - \zeta_i A(|\vec{\zeta}|) \partial_i \tau_{\mathrm{H}0} - \frac{1}{2} \zeta_i \zeta_j B(|\vec{\zeta}|) (\partial_j u_{i\mathrm{H}0} + \partial_i u_{j\mathrm{H}0}),
\label{e:phi_H1}
\\
& \phi_{\mathrm{K}1} = \varphi_1^{(0)}(\eta,\zeta_1,|\overline{\vec{\zeta}}|) \,(\partial_1 \tau_{\mathrm{H}0})_{0}
\nonumber \\
& \qquad 
+ \overline{\zeta}_i \left[ \varphi_1^{(1)}(\eta,\zeta_1,|\overline{\vec{\zeta}}|) \,n_j (\partial_j u_{i\mathrm{H}0} + \partial_i u_{j\mathrm{H}0})_0 
\right.
\nonumber \\
& \qquad \qquad \left. + \varphi_2^{(1)}(\eta,\zeta_1,|\overline{\vec{\zeta}}|) \,(\partial_i \tau_{\mathrm{H}0})_0
\right], 
\quad \eta = \frac{x_1 + \frac{\pi}{2}}{\varepsilon}.
\label{e:phi_K1}
\end{align}
\end{subequations}
Here,
\begin{enumerate}
\item $\phi_{\mathrm{eH}m}$ is a linear combination of $(1,\zeta_i,|\vec{\zeta}|)$ forming the (linearized) local Maxwellian
    \begin{eqnarray*}
    \phi_{\mathrm{eH}m} = P_{\mathrm{H}m} + 2 \zeta_i u_{i\mathrm{H}m} + \left(|\vec{\zeta}|^2 - \frac{5}{2}\right) \tau_{\mathrm{H}m},
\quad m=0,1.
    \end{eqnarray*}
\item The functions $A(|\vec{\zeta}|)$ and $B(|\vec{\zeta}|)$ are the solutions to the integral equations
\begin{eqnarray*}
&& \mathscr{L}(\zeta_i A) = - \zeta_i \left(|\vec{\zeta}|^2 - \frac{5}{2}\right), \quad \text{with} \ \ \langle |\vec{\zeta}|^2 A\rangle = 0,
\\
&& \mathscr{L}(\zeta_{ij} B) = - 2 \zeta_{ij},
\end{eqnarray*}
where $\zeta_{ij} = \zeta_i \zeta_j - \frac{|\vec{\zeta}|^2}{3} \delta_{ij}$.
\item $\eta$ is a stretched coordinate of $x_1$ near the boundary $x_1=-\frac{\pi}{2}$, adequate to describe the Knudsen-layer corrections.
\item $\overline{\vec{\zeta}}$ is a projection of $\vec{\zeta}$ onto a plane orthogonal to $\vec{n}=(1,0,0)$, i.e.,
\begin{equation*}
\overline{\zeta}_i = \zeta_j (\delta_{ij} - n_i n_j).
\end{equation*}
%
\item The symbol $(\cdot)_0$ indicates the value on $x_1=-\frac{\pi}{2}$.
\item The functions $\varphi_1^{(0)} = \varphi_1^{(0)}(\eta,\zeta_1,|\overline{\vec{\zeta}}|)$ and $\varphi_j^{(1)} = \varphi_j^{(1)}(\eta,\zeta_1,|\overline{\vec{\zeta}}|)$, $j = 1, 2$, solve the following half-space problems (Knudsen-layer problems):
\begin{subequations}
\begin{align}
& \zeta_1 \partial_\eta \varphi_1^{(0)} = \mathscr{L}(\varphi_1^{(0)}),
\\
& \varphi_1^{(0)} = - (|\vec{\zeta}|^2 - 2) c_1^{(0)} + \zeta_1 A(|\vec{\zeta}|) 
\nonumber \\
& \qquad \
+ 4 \int_0^\infty \int_{-\infty}^0 |\zeta_1| |\overline{\vec{\zeta}}| \varphi_1^{(0)} \E^{-|\vec{\zeta}|^2} \D\zeta_1 \D |\overline{\vec{\zeta}}|,
\quad \zeta_1 > 0, \quad \eta = 0,
\\
& \varphi_1^{(0)} \to 0, \quad \text{as} \ \  \eta \to \infty;
\end{align}
\end{subequations}
\begin{subequations}
\begin{align}
& \zeta_1 \partial_\eta \varphi_j^{(1)} = \mathscr{L}(\varphi_j^{(1)}),
\quad j \in \{1,2\},
\\
& \varphi_j^{(1)} = - 2 b_j^{(1)} + J_j, \quad \zeta_1 > 0, \quad \eta = 0,
\\
& \varphi_j^{(1)} \to 0, \quad \text{as} \ \ \eta \to \infty,
\end{align}
\end{subequations}
with 
\begin{subequations}
\begin{align}
& J_1 = \zeta_1 B(|\vec{\zeta}|), \quad J_2 = A(|\vec{\zeta}|),
\\
& c_1^{(0)}, \quad b_j^{(1)} \quad (j=1,2): \quad \text{constants}.
\end{align}
\end{subequations}
Note that $|\vec{\zeta}| = \sqrt{\zeta_1^2 + |\overline{\vec{\zeta}}|^2}$.
It is known that there exists a solution to the problem if and only if the constant $c_1^{(0)}$ or $b_j^{(0)}$ takes a special value and that the solution is unique \cite{Bardos-Caflisch-Nicolaenko_CPA_1986,Coron-Golse-Sulem_CPA_1988,Sone02}. It has also been proved that the solution decays exponentially fast as $\eta \to \infty$.
\end{enumerate}
Suppose that the functions $A$, $B$, $\varphi_1^{(0)}$, and $\varphi_i^{(1)}$, $i=1,2$, are known.
Then, the functional dependency of $\phi_{\mathrm{H}m}$ and $\phi_{\mathrm{K}m}$ on the molecular velocity $\vec{\zeta}$ is prescribed through these auxiliary functions and  $\phi_{\mathrm{eH}m}$.
On the other hand, the spatial dependency enters through those of $u_{i\mathrm{H}m}(\vec{x})$, $\tau_{\mathrm{H}m}(\vec{x})$, and $P_{\mathrm{H}m}(\vec{x})$ (and their spatial derivatives when $m \ge 1$). The dependency of $u_{i\mathrm{H}m}$, $\tau_{\mathrm{H}m}$, and $P_{\mathrm{H}m}$, and $\omega_{\mathrm{H}m}$ on $\vec{x}$ are obtained via the fluid-dynamic-type problems stated next.

\medskip
\noindent
\textbf{Stokes problem.}
The expansion coefficients of the macroscopic quantities $h_{\mathrm{H}m}$ ($h=\omega,\,u_i,\,\tau,\,P$) are described by the following equations and boundary conditions on $x_1=-\frac{\pi}{2}$. The equations are
\begin{align}
& \partial_i P_{\mathrm{H}0} = 0, \label{e:fde_P0}
\end{align}
\begin{subequations}
\begin{align}
& \partial_i u_{i\mathrm{H}m} = 0, && (\text{continuity equation})
\label{e:fde_continuity}
\\
& \gamma_1 \Delta u_{i\mathrm{H}m} - \partial_i P_{\mathrm{H}m+1} = 0,
&& (\text{equation of motion})
\label{e:fde_velo}
\\
& \Delta \tau_{\mathrm{H}m} = 0,
&& (\text{energy equation})
\label{e:fde_tau}
\\
& \omega_{\mathrm{H}m} = P_{\mathrm{H}m} - \tau_{\mathrm{H}m},
&& (\text{equation of state})
\label{e:fde_state}
\end{align}
\end{subequations}
($m=0,1,\ldots$). The boundary conditions on $x_1 = -\frac{\pi}{2}$ are
\begin{subequations}
\begin{align} \text{Order $\varepsilon^0$:} \quad 
& u_{1\mathrm{H}0} = u_{2\mathrm{H}0} = u_{3\mathrm{H}0} = 0, \quad \tau_{\mathrm{H}0} = \oltauw,
\label{e:bc_noslip/nojump}
\\
\text{Order $\varepsilon^1$:} \quad 
& u_{1\mathrm{H}1} = 0, \quad \tau_{\mathrm{H}1} = c_1^{(0)} \partial_1 \tau_{\mathrm{H}0},
\label{e:bc_slip_1}\\
& u_{j\mathrm{H}1} t_j = b_1^{(1)} t_j n_k (\partial_j u_{k\mathrm{H}0} + \partial_k u_{j\mathrm{H}0}) + b_2^{(1)} t_j \partial_j \tau_{\mathrm{H}0}.
\label{e:bc_slip_2}
\end{align}
\end{subequations}
Here, $\Delta = \partial_1^2 + \partial_2^2 + \partial_3^2$ is the Laplacian,
the \textit{viscosity} $\gamma_1>0$ is defined by
\begin{align}
\gamma_1 =\frac{2}{15}\langle |\vec{\zeta}|^4 B \rangle,
\end{align}
$t_i$ is any unit vector orthogonal to $\vec{n}=(1,0,0)$,
and $b_i^{(1)}$ ($i=1,2$) and $c_1^{(0)}$, known as the slip/jump coefficients, are the same constants arising in the Knudsen-layer problem introduced above.
The numerical value of $\gamma_1$ and those of the slip/jump coefficients for a hard-sphere gas are obtained as $\gamma_1=1.270042427$ and
$(b_1^{(1)},b_2^{(1)},c_1^{(0)})=(-k_0,-K_1,d_1)=(1.2540,0.6465,2.4001)$, where $k_0$, $K_1$, and $d_1$ are the notations used in \cite{Sone02,Sone07}.

It should be noted that, since we are seeking a solution that is symmetric with respect to $x_1=0$, the above system should be supplemented by an appropriate reflection condition at $x_1 = 0$. A similar comment applies throughout the paper and will not be repeated in the sequel.


\medskip
\noindent
\textbf{Solution procedure.}
For a given $\oltauw$, the process to obtain the solution $\phi$ to the order $\varepsilon$ is as follows:
\begin{enumerate}
\item From Eq.~\eqref{e:fde_P0}, $P_{\mathrm{H}0} = p_0$ (constant). 
\item Solve Eqs.~\eqref{e:fde_continuity}--\eqref{e:fde_tau} for $m=0$ under the condition \eqref{e:bc_noslip/nojump} to obtain $u_{\mathrm{H}0}$, $P_{\mathrm{H}1}$, and $\tau_{\mathrm{H}0}$.
Note that $P_{\mathrm{H}1}$ is determined up to an additive constant (say, $p_1$). 
Compute $\omega_{\mathrm{H}0}$ from Eq.~\eqref{e:fde_state} with $m=0$. 
The leading-order solution $\phi_{\mathrm{H}0}$ is derived from Eq.~\eqref{e:phi_H0}.
\item 
Solve Eqs.~\eqref{e:fde_continuity}--\eqref{e:fde_tau} for $m=1$ under the conditions~\eqref{e:bc_slip_1} and \eqref{e:bc_slip_2} to obtain $u_{\mathrm{H}1}$, $P_{\mathrm{H}2}$, and $\tau_{\mathrm{H}1}$.
Note that $P_{\mathrm{H}2}$ is determined up to an additive constant (say, $p_2$). 
Compute $\omega_{\mathrm{H}1}$ from Eq.~\eqref{e:fde_state} with $m=1$. 
The first order solution $\phi_{\mathrm{H}1}+\phi_{\mathrm{K}1}$ is obtained from Eqs.~\eqref{e:phi_H1} and \eqref{e:phi_K1}.
\end{enumerate}
In the above procedure, $P_{\mathrm{H}m}$, $\omega_{\mathrm{H}m}$, and $\phi_{\mathrm{H}m}$ are determined up to a (common) additive constant $p_m$ at each $m$, although $u_{i\mathrm{H}m}$ and $\tau_{\mathrm{H}m}$ are determined without such ambiguities. A physical argument can single out a solution. For example, we can specify the gas pressure at a certain point in the domain or specify the average gas density in the whole domain. Another possibility to remove the ambiguity might be through a symmetry argument (depending on $\oltauw$), as in the next section.

\section{\label{sec:discontinuou}Case of a discontinuous wall temperature}

Now we return to the original problem. Again, we assume that the solution is symmetric with respect to $x_1=0$
and restrict the domain in $D^-$. Moreover, we seek the solution that is antisymmetric with respect to $x_2=0$, i.e.,
\begin{align}
\phi(x_1,-x_2,x_3,\zeta_1,-\zeta_2,\zeta_3) = - \phi(x_1,x_2,x_3,\zeta_1,\zeta_2,\zeta_3).
\end{align}
Henceforth, we assume that the solution is $x_3$-independent, i.e., $\partial_3=0$, and even in $\zeta_3$ (hence, $u_3 = 0$).

First, leaving aside the fact that the boundary condition is discontinuous at $(x_1,x_2)=(-\frac{\pi}{2},0)$, we look for a solution to the system \eqref{e:LB_2}--\eqref{e:diffuse_2} in the form
\begin{equation}
\phi = \phi_{\mathrm{HK}} = \phi_{\mathrm H} + \phi_{\mathrm K}.
\end{equation}
Here, $\phi_{\mathrm{H}}$ is the Hilbert solution, $\phi_{\mathrm{K}}$ the Knudsen-layer correction, and $\phi_{\mathrm{HK}}$ their sum. Hereafter, we call $\phi_{\mathrm{HK}}$ the Hilbert-Knudsen (HK) solution. Note that $\phi_{\mathrm{H}}$ and $\phi_{\mathrm{K}}$ are subject to the conditions
\begin{align}
\partial_i \phi_{\mathrm{H}} = O(\phi_{\mathrm{H}}), \quad i=1,2,
\quad 
\partial_1 \phi_{\mathrm{K}} = O(\phi_{\mathrm{K}}/\varepsilon),
\quad
\partial_2 \phi_{\mathrm{K}} = O(\phi_{\mathrm{K}}).
\end{align}
As in the previous section, $\phi_{\mathrm{H}}$ and $\phi_{\mathrm{K}}$, and thus $\phi_{\mathrm{HK}}$, are expanded in $\varepsilon$ as
\begin{subequations}
\begin{align}
& \phi_{\mathrm{H}} = \phi_{\mathrm{H}0} + \varepsilon \phi_{\mathrm{H}1} + \cdots,
\\
& \phi_{\mathrm{K}} = \varepsilon \phi_{\mathrm{K}1} + \cdots,
\\
& \phi_{\mathrm{HK}} = \phi_{\mathrm{HK}0} + \varepsilon \phi_{\mathrm{HK}1} + \cdots,
\end{align}
\end{subequations}
with 
\begin{equation}
 \phi_{\mathrm{HK}0} =  \phi_{\mathrm{H}0}, \quad  \phi_{\mathrm{HK}1} = \phi_{\mathrm{H}1} + \phi_{\mathrm{K}1}.
\end{equation}
To obtain $\phi_{\mathrm{HK}0}$ and $\phi_{\mathrm{HK}1}$, We apply the solution algorithm given in the previous section.

\medskip
\noindent \textbf{Step 1.} 
The leading-order pressure is $P_{\mathrm{H}0}=p_0$ (constant).
We chose $P_{\mathrm{H}0}=p_0=0$ in view of the antisymmetry of the solution.

\medskip
\noindent \textbf{Step 2.} 
The Stokes problem to determine $u_{i\mathrm{H}0}$ and $\tau_{\mathrm{H}0}$ reads
\begin{subequations}
\begin{align}
& \partial_i u_{i\mathrm{H}0} = 0, \quad \gamma_1 \Delta u_{i\mathrm{H}0} - \partial_i P_{\mathrm{H}1}= 0,
\quad
\Delta \tau_{\mathrm{H}0} = 0,
\quad
\omega_{\mathrm{H}0} = - \tau_{\mathrm{H}0},
\quad \text{in} \ \ D^-, \label{e:stokes_prob_order0}
\\
& u_{i\mathrm{H}0} = 0, \quad \tau_{\mathrm{H}0} = \pm \tauw, \quad \text{on} \ \ x_1 = -\frac{\pi}{2}, \ \  x_2 \gtrless 0.
\end{align}
\end{subequations}
The solution is given by
\begin{subequations}
\begin{align}
& u_{i\mathrm{H}0} = 0, \quad P_{\mathrm{H}1} = 0, 
\\
& \tau_{\mathrm{H}0} = - \omega_{\mathrm{H}0} = \frac{\tauw}{\pi} \text{Arg}\left(\frac{1+\sin z}{1-\sin z}\right),
\quad z = x_1 + \imag \,x_2,
\end{align}
\end{subequations}
where $\imag$ is the imaginary unit, and the additive constant in $P_{\mathrm{H}1}$ is chosen to be zero because of the solution's antisymmetry.
Hence, we obtain the leading-order HK solution as
\begin{align}
\phi_{\mathrm{HK}0} = \phi_{\mathrm{H}0} = \left(|\vec{\zeta}|^2 - \frac{5}{2} \right) \tau_{\mathrm{H}0}
= \left(|\vec{\zeta}|^2 - \frac{5}{2} \right) \frac{\tauw}{\pi} \text{Arg}\left(\frac{1+\sin z}{1-\sin z}\right).
\end{align}

\medskip
\noindent \textbf{Step 3.} 
The Stokes problem for the first order in $\varepsilon$ is reduced to
\begin{subequations}
\begin{align}
& \partial_i u_{i\mathrm{H}1} = 0, \quad \gamma_1 \Delta u_{i\mathrm{H}1} - \partial_i P_{\mathrm{H}2} = 0,
\quad
\Delta \tau_{\mathrm{H}1} = 0,
\quad
\omega_{\mathrm{H}1} = - \tau_{\mathrm{H}1},
\quad \text{in} \ \ D^-,
\label{e:stokes_prob_order1}
\\
& u_{i\mathrm{H}1} = 0, \quad \tau_{\mathrm{H}1} = - \frac{2\tauw c_1^{(0)}}{\pi} \frac{1}{\sinh x_2},
\quad \text{on} \ \ x_1 = - \frac{\pi}{2}, \ \ x_2 \ne 0.
\label{e:stokes_prob_bc_1}
\end{align}
\end{subequations}
The solution is given by
\begin{subequations}
\begin{align}
& u_{i\mathrm{H}1} = 0, \quad P_{\mathrm{H}2} = 0,
\\
& \tau_{\mathrm{H}1} = - \omega_{\mathrm{H}1} = -\frac{8 \tauw c_1^{(0)}}{\pi^2} \frac{x_2 \cos x_1 \cosh x_2 + x_1 \sin x_1 \sinh x_2}{\cos (2x_1) + \cosh (2x_2)},
\label{e:sol_tau_H1}
\end{align}
\end{subequations}
where the additive constant in $P_{\mathrm{H}2}$ is chosen to be zero because of the solution's antisymmetry.
Hence, we obtain the first-order HK solution $\phi_{\mathrm{HK1}}$ as
\begin{subequations}
\begin{align}
\phi_{\mathrm{H}1} & = \left(|\vec{\zeta}|^2 - \frac{5}{2} \right) \tau_{\mathrm{H}1} - \zeta_i A(|\vec{\zeta}|) \partial_i \tau_{\mathrm{H}0}
\nonumber \\
& = - \frac{8 \tauw c_1^{(0)}}{\pi^2} \left(|\vec{\zeta}|^2 - \frac{5}{2} \right) \frac{x_2 \cos x_1 \cosh x_2 + x_1 \sin x_1 \sinh x_2}{\cos (2x_1) + \cosh (2x_2)}
\nonumber \\
& \quad \!
- \frac{4 \tauw}{\pi} A(|\vec{\zeta}|) \frac{\zeta_1 \sin x_1 \sinh x_2 + \zeta_2 \cos x_1 \cosh x_2}{\cos (2x_1) + \cosh (2 x_2)},
\label{e:Hilbert_order1}
\\
\phi_{\mathrm{K}1} & = - \frac{2 \tauw}{\pi} \frac{1}{\sinh x_2} \varphi_1^{(0)}\!\left(\frac{x_1 + \frac{\pi}{2}}{\varepsilon},\zeta_1,|\overline{\vec{\zeta}}|\right), 
\label{e:KL_order1}
\\
\phi_{\mathrm{HK1}} & = \phi_{\mathrm{H1}} + \phi_{\mathrm{K1}}.
\end{align}
\end{subequations}

\medskip\noindent
\textbf{Drawbacks.}
We have obtained the first two terms of the HK solution $\phi_{\mathrm{HK}} = \phi_{\mathrm{HK}0} + \varepsilon \phi_{\mathrm{HK}1}$ disregarding the fact that the boundary data is discontinuous at $(x_1,x_2)=(-\frac{\pi}{2},0)$. This solution has the following drawbacks.
\begin{enumerate}
\item The solution does not produce any non-zero flow velocity, which is not meaningful. Note that a non-uniform surface temperature of a body usually causes a rarefied gas flow such as the thermal creep. This remains true even if the temperature distribution is piecewise uniform with a jump discontinuity \cite{Aoki-Takata-Aikawa-Golse_PHF97}.
\item Near the point $(x_1,x_2)=(-\frac{\pi}{2},0)$, the $\phi_{\mathrm{HK}0}$ and $\phi_{\mathrm{HK}1}$ have the following asymptotic properties:
\begin{subequations}
\begin{align}
\phi_{\mathrm{HK}0} =&\,\tauw \left(|\vec{\zeta}|^2 - \frac{5}{2} \right) \left(\frac{2}{\pi} \theta + \frac{r^2}{6\pi} \sin (2\theta) + O(r^4) \right),
\label{e:phi_HK0_expansion}
\\
\phi_{\mathrm{HK}1} =& - \frac{2 \tauw}{\pi} \left[\frac{c_1^{(0)} \sin \theta}{r} \left(|\vec{\zeta}|^2 - \frac{5}{2} \right) + \frac{\zeta_\theta}{r} A(|\vec{\zeta}|) + \frac{1}{x_2} \varphi_1^{(0)}\!\left(\frac{x_1 + \frac{\pi}{2}}{\varepsilon},\zeta_1,|\overline{\vec{\zeta}}|\right) \right] 
\nonumber \\
&+ O(r),
\label{e:phi_HK1_expansion}
\end{align}
\end{subequations}
as $r \searrow 0$, where
$$
r = \sqrt{\left(x_1 + \frac{\pi}{2}\right)^2 + x_2^2}, \quad 
\theta = \text{Arctan}\left(\frac{x_2}{x_1 + \frac{\pi}{2}}\right),
$$
and $\zeta_\theta = - \zeta_1 \sin \theta + \zeta_2 \cos \theta$.
Thus, $|\phi_{\mathrm{HK}1}|$ grows indefinitely with the rate $r^{-1}$ as $r \searrow 0$. In other words, the $\varepsilon$-expansion of $\phi_{\mathrm{HK}}$ is meaningful only in the region $r \gg \varepsilon$ in $D^-$.
\end{enumerate}

\subsection{Knudsen zone} Motivated by the above observation, we now look for a solution in the form
\begin{align}
& \phi = 
\begin{cases}
\phi_{\mathrm{HK}} = \phi_{\mathrm{H}} + \phi_{\mathrm{K}} & \text{in} \ \ D^- \cap \{(x_1,x_2)\ | \ r \gg \varepsilon, \ r= \sqrt{\left(x_1+\frac{\pi}{2}\right)^2 + x_2^2} \},\\[2mm]
\phi_{\mathrm{Z}} & \text{in} \ \ D^- \cap \{(x_1,x_2)\ | \ r \ll 1, \ r= \sqrt{\left(x_1+\frac{\pi}{2}\right)^2 + x_2^2} \},
\end{cases}
\end{align}
allowing $\phi_{\mathrm{HK}}$ and $\phi_{\mathrm{Z}}$ to overlap in the region $\varepsilon \ll r \ll 1$.
Here, $\phi_{\mathrm{Z}}$ replaces $\phi_{\mathrm{HK}}$ in the region close to the point of discontinuity $(x_1,x_2)=(-\frac{\pi}{2},0)$ (i.e., the Knudsen zone). In the Knudsen zone, the length scale of variation of $\phi_{\mathrm{Z}}$ is assumed to be of the order of $\varepsilon$, i.e., $\partial_i \phi_{\mathrm{Z}} = O(\phi_{\mathrm{Z}}/\varepsilon)$ ($i=1,2$).

To analyze $\phi_{\mathrm{Z}}$, we introduce new spatial variables by
\begin{align}
x_i = - \frac{\pi}{2} \delta_{i1} + \varepsilon y_i, \quad i = 1,2,
\end{align}
and assume that $\phi_{\mathrm{Z}} = \phi_{\mathrm{Z}}(y_1,y_2,\vec{\zeta})$.
Expanding $\phi_{\mathrm{Z}}$ in the form
\begin{align}
& \phi_{\mathrm{Z}} = \phi_{\mathrm{Z}0} + \varepsilon \phi_{\mathrm{Z}1} + \cdots,
\end{align}
the zeroth-order term $\phi_{\mathrm{Z}0}$ satisfies the following equation and boundary conditions:
\begin{subequations}\label{e:KnZone}
\begin{align}
& \zeta_1 \frac{\partial \phi_{\mathrm{Z0}}}{\partial y_1} + \zeta_2 \frac{\partial \phi_{\mathrm{Z0}}}{\partial y_2}
= \mathscr{L}(\phi_{\mathrm{Z0}}), \qquad (y_1 > 0, \ -\infty < y_2 < \infty),
\label{e:eq_phiZ0}
\\
& \phi_{\mathrm{Z0}} = 2 \sqrt{\pi} \int_{\zeta_1 < 0} |\zeta_1| \phi_{\mathrm{Z0}} E \pm (|\vec{\zeta}|^2 - 2) \tauw,
\quad \zeta_1 > 0, \qquad (y_1=0, \ y_2 \gtrless 0),
\label{e:bc_phi_Z0}
\\
& \phi_{\mathrm{Z0}} \to \frac{2 \tauw \Gamma_z^{(1)}}{|\vec{y}|} \zeta_r \sin (2\theta) + \frac{2 \tauw}{\pi} \left(|\vec{\zeta}|^2 - \frac{5}{2} \right) \left(\theta - \frac{c_1^{(0)}}{|\vec{y}|} \sin \theta \right)
\nonumber \\
& \qquad \quad - \frac{2 \tauw}{\pi} \left(\frac{\zeta_\theta}{|\vec{y}|} A(|\vec{\zeta}|) + \frac{1}{y_2} \varphi_1^{(0)}(y_1,\zeta_1,|\overline{\vec{\zeta}}|) \right), \quad \text{as} \ \ |\vec{y}| \to \infty,
\label{e:mc_phi_Z0} \\
& \theta = \text{Arctan}\left(\frac{y_2}{y_1}\right), \quad \zeta_r = \zeta_1 \cos \theta + \zeta_2 \sin \theta,
\quad \zeta_\theta = - \zeta_1 \sin \theta + \zeta_2 \cos \theta,
\end{align}
\end{subequations}
where $\Gamma_z^{(1)}$ is a constant that represents the far-field asymptotic property of $\phi_{\mathrm{Z0}}$, and should be determined together with the solution. This problem can be viewed as a two-dimensional analog of the thermal creep flow \cite{Sone66,Loyalka_PHF1971,Ohwada-Sone-Aoki89}, and represents a ``reaction'' of a rarefied gas to a forced temperature variation in the gas.
We give further details on the derivation of \eqref{e:mc_phi_Z0} in Appendix.

\subsection{A source-sink condition for the flow velocity}
Let us assume that $\phi_{\mathrm{Z}0}$ is known including $\Gamma_{z}^{(1)}$.
We consider a point in $D^-$ such that $\varepsilon \ll r = \sqrt{(x_1 + \frac{\pi}{2})^2 + x_2^2} \ll 1$,
and consider the asymptotic behavior of $\phi_{\mathrm{Z}}$ in the limit $\varepsilon \searrow 0$, keeping $r \,(= \varepsilon |\vec{y}|)$ fixed. With the aid of \eqref{e:mc_phi_Z0}, this is obtained as
\begin{align}
\phi_{\mathrm{Z}} & = \varepsilon \frac{2 \tauw \Gamma_{z}^{(1)}}{r} \zeta_r \sin (2\theta) + \frac{2 \tauw}{\pi} \left(|\vec{\zeta}|^2 - \frac{5}{2} \right) \left(\theta - \varepsilon \frac{c_1^{(0)}}{r} \sin \theta \right)
\nonumber \\
& \quad - \varepsilon \frac{2 \tauw}{\pi} \left(\frac{\zeta_\theta}{r} A(|\vec{\zeta}|) + \frac{1}{x_2} \varphi_1^{(0)}\!\left(\frac{x_1 + \frac{\pi}{2}}{\varepsilon},\zeta_1,|\overline{\vec{\zeta}}|\right) \right)
\nonumber \\
& = \frac{2 \tauw}{\pi} \left(|\vec{\zeta}|^2 - \frac{5}{2} \right) \theta + \varepsilon \bigg[
\frac{2 \tauw \Gamma_{z}^{(1)}}{r} \zeta_r \sin (2\theta) - \frac{2 \tauw}{\pi} \left(|\vec{\zeta}|^2 - \frac{5}{2} \right) \frac{c_1^{(0)}}{r} \sin \theta
\nonumber \\
& \quad - \frac{2 \tauw}{\pi} \frac{\zeta_\theta}{r} A(|\vec{\zeta}|) - \frac{2 \tauw}{\pi} \frac{1}{x_2} \varphi_1^{(0)}\!\left(\frac{x_1 + \frac{\pi}{2}}{\varepsilon},\zeta_1,|\overline{\vec{\zeta}}|\right) \bigg],
\quad \text{as} \ \ \varepsilon \searrow 0 \ \  \text{with} \ \  r \ \ \text{fixed},
\end{align}
where $\theta = \text{Arctan}(\frac{x_2}{x_1 + \frac{\pi}{2}})$.
Hence, $\phi_{\mathrm{HK}}$ is matched to the first two terms of $\phi_{\mathrm{Z}}$ if 
\begin{align}
\phi_{\mathrm{HK}1} & \to \frac{2 \tauw \Gamma_{z}^{(1)}}{r} \zeta_r \sin (2\theta) - \frac{2 \tauw}{\pi} \left(|\vec{\zeta}|^2 - \frac{5}{2} \right) \frac{c_1^{(0)}}{r} \sin \theta
\nonumber \\
& \quad - \frac{2 \tauw}{\pi} \frac{\zeta_\theta}{r} A(|\vec{\zeta}|) - \frac{2 \tauw}{\pi} \frac{1}{x_2} \varphi_1^{(0)}\!\left(\frac{x_1 + \frac{\pi}{2}}{\varepsilon},\zeta_1,|\overline{\vec{\zeta}}|\right),
\quad \text{as} \ \ r \to 0.
\end{align}
Separating the Hilbert part from the Knudsen-layer part, we have
\begin{align}
\phi_{\mathrm{H}1} & \to \frac{2 \tauw \Gamma_{z}^{(1)}}{r} \zeta_r \sin (2\theta) - \frac{2 \tauw}{\pi} \left(|\vec{\zeta}|^2 - \frac{5}{2} \right) \frac{c_1^{(0)}}{r} \sin \theta
- \frac{2 \tauw}{\pi} \frac{\zeta_\theta}{r} A(|\vec{\zeta}|),
\end{align}
as $r \to 0$. Thus, the radial and circumferential components of the flow velocity $u_{r\mathrm{H}1} = \langle \zeta_r \phi_{\mathrm{H}1} \rangle$ and $u_{\theta\mathrm{H}1} = \langle \zeta_\theta \phi_{\mathrm{H}1} \rangle$ near the point of discontinuity behave as
\begin{align}
u_{r\mathrm{H}1} \to \frac{\tauw \Gamma_{z}^{(1)}}{r} \sin (2\theta), \quad 
u_{\theta \mathrm{H}1} \to 0,
\quad \text{as} \ \ r \to 0,
\label{e:source-sink}
\end{align}
with
\begin{align}
r = \sqrt{\left( x_1 + \frac{\pi}{2} \right)^2 + x_2^2}, \quad \theta = \text{Arctan}\left(\frac{x_2}{x_1 + \frac{\pi}{2}}\right).
\end{align}
The condition describes a source-sink pair located at $(x_1,x_2)=(-\frac{\pi}{2},0)$ and serves as a ``boundary condition'' that provokes a non-vanishing flow velocity in the Stokes system. As we will see later (Sect.~\ref{sec:numerical}), $\Gamma_z^{(1)}$ is likely to be a positive number. Thus, a sink flow toward the discontinuity point appears in the region $x_2<0$ and a source flow in the region $x_2>0$. 

To summarize, after the consideration of the Knudsen zone, \textbf{Step 3} should be replaced by

\medskip
\noindent\textbf{Step 3'.}
The Stokes problem for the first order in $\varepsilon$ is given by
\begin{subequations}
\begin{align}
& \partial_i u_{i\mathrm{H}1} = 0, \quad \gamma_1 \Delta u_{i\mathrm{H}1} - \partial_i P_{\mathrm{H}2} = 0,
\quad
\Delta \tau_{\mathrm{H}1} = 0,
\quad
\omega_{\mathrm{H}1} = - \tau_{\mathrm{H}1},
\quad \text{in} \ \ D^-,
\\
& u_{i\mathrm{H}1} = 0, \quad \tau_{\mathrm{H}1} = - \frac{2\tauw c_1^{(0)}}{\pi} \frac{1}{\sinh x_2},
\quad \text{on} \ \ x_1 = - \frac{\pi}{2}, \ \ x_2 \ne 0,
\\
& u_{r\mathrm{H}1} \to \frac{\tauw \Gamma_{z}^{(1)}}{r} \sin (2\theta), \quad 
u_{\theta \mathrm{H}1} \to 0,
\quad \text{as} \ \ r = \sqrt{\left(x_1 + \frac{\pi}{2}\right)^2 + x_2^2} \to 0.
\label{e:source-sink_2}
\end{align}
\end{subequations}
The solution $\tau_{\mathrm{H}1}$ is given by \eqref{e:sol_tau_H1}, while $(u_{1\mathrm{H}1},u_{2\mathrm{H}1})$ can be obtained, for instance, by applying the Fourier transform. With these solutions,
the first-order HK solution $\phi_{\mathrm{HK1}}$ is given by
\begin{subequations}
\begin{align}
\phi_{\mathrm{H}1} 
&= 2 \zeta_1 u_{1\mathrm{H}1} + 2 \zeta_2 u_{2\mathrm{H}1} 
\nonumber \\
& - \frac{8 \tauw c_1^{(0)}}{\pi^2} \left(|\vec{\zeta}|^2 - \frac{5}{2} \right) \frac{x_2 \cos x_1 \cosh x_2 + x_1 \sin x_1 \sinh x_2}{\cos (2x_1) + \cosh (2x_2)}
\nonumber \\
& - \frac{4 \tauw}{\pi} A(|\vec{\zeta}|) \frac{\zeta_1 \sin x_1 \sinh x_2 + \zeta_2 \cos x_1 \cosh x_2}{\cos (2x_1) + \cosh (2 x_2)},
\\
\phi_{\mathrm{K}1} & = - \frac{2 \tauw}{\pi} \frac{1}{\sinh x_2} \varphi_1^{(0)}\!\left(\frac{x_1 + \frac{\pi}{2}}{\varepsilon},\zeta_1,|\overline{\vec{\zeta}}|\right), 
\\
\phi_{\mathrm{HK1}} & = \phi_{\mathrm{H1}} + \phi_{\mathrm{K1}}.
\end{align}
\end{subequations}
Note that $\phi_{\mathrm{K}1}$ has not been modified from \eqref{e:KL_order1}.


\section{\label{sec:numerical}Numerical results for the Knudsen-zone problem}
\begin{figure}
    \centering
    \subfigure{%
    \includegraphics[width=0.5\columnwidth]{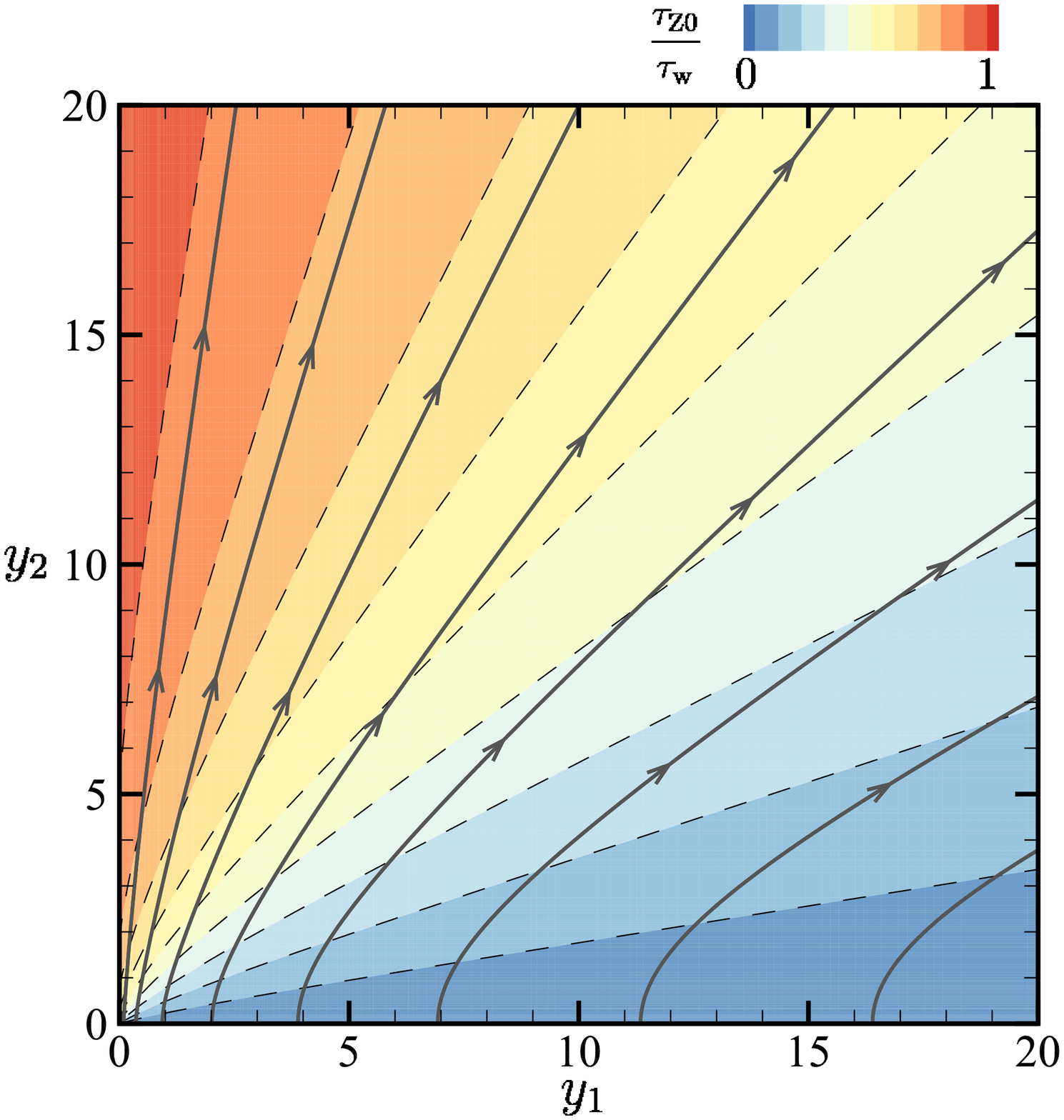}}%
    \subfigure{%
    \includegraphics[width=0.5\columnwidth]{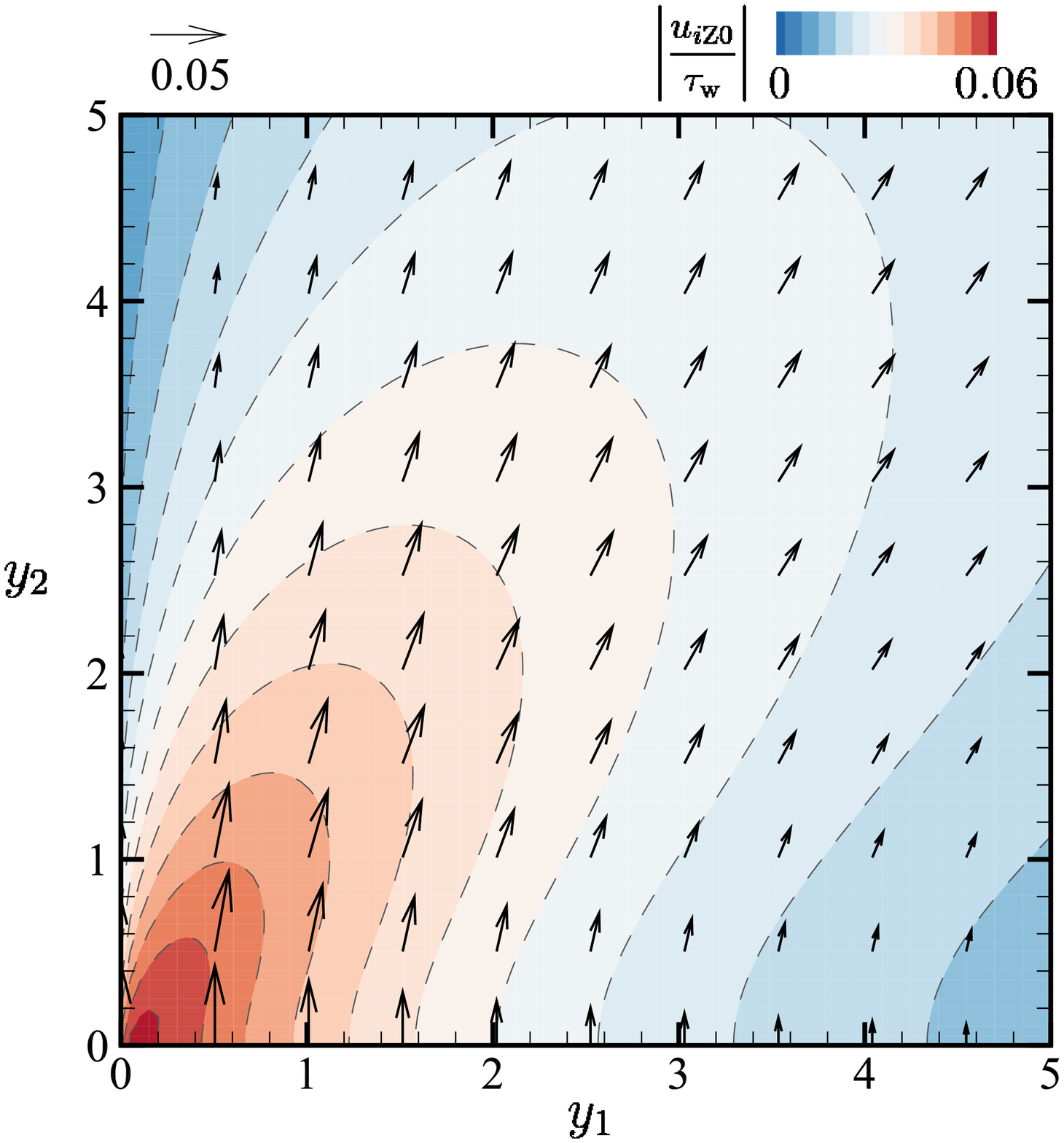}}%
    \caption{Numerical results for the Knudsen-zone problem based on the (linearized) BGK collision operator. (a) The thick gray curves with arrows show the streamlines of the flow velocity $\tauw^{-1}(u_{1\mathrm{Z}0},u_{2\mathrm{Z}0})$, and the dashed contours show the temperature $\tau_{\mathrm{Z}0}/\tauw$. (b) A magnified figure near the origin. The arrow indicates the flow-velocity vector $\tauw^{-1}(u_{1\mathrm{Z}0},u_{2\mathrm{Z}0})$ at its starting point, and the contours visualize the absolute value.}
    \label{fig:flow}
\end{figure}

Finally, we show some preliminary results for the Knudsen-zone problem. To simplify the numerical analysis, we employ the Bhatnagar-Gross-Krook (BGK) collision operator \cite{Bhatnagar-Gross-Krook54,Welander54} instead of the Boltzmann collision operator.
The linearized BGK collision operator is well-known and its explicit form is omitted \cite{Sone07}.
Figure~\ref{fig:flow}(a) shows the streamlines of the flow velocity $(u_{1\mathrm{Z}0},u_{2\mathrm{Z}0})$ and the (perturbed) temperature $\tau_{\mathrm{Z}0}$ in the upper-half domain $y_1 \ge 0$ and $y_2 \ge 0$. Here, $u_{i\mathrm{Z}0}$ and $\tau_{\mathrm{Z}0}$ are defined by
\begin{align}
u_{i\mathrm{Z}0} = \langle \zeta_i \phi_{\mathrm{Z}0} \rangle, \quad i=1,2, \quad \tau_{\mathrm{Z}0} = \frac{2}{3} \left \langle \left(|\vec{\zeta}|^2 - \frac{3}{2} \right) \phi_{\mathrm{Z}0} \right \rangle.
\end{align}
Note that the wall temperature is discontinuous at $y_2=0$ along $y_1=0$ (the plates' temperature is $T_0 (1 \pm \tauw)$ for $y_2 \gtrless 0$). 
Figure~\ref{fig:flow}(b) shows the flow-velocity vector $(u_{1\mathrm{Z}0},u_{2\mathrm{Z}0})$ and its absolute value near the origin.
As seen from these figures, a flow is induced in the positive $y_2$ direction, which exhibits a diverging flow pattern in the region far from the origin. Note that, by the antisymmetry, it implies that there is a shrinking flow toward the origin in the region $y_2 < 0$. The flow speed is strongest near the discontinuity point and decreases as $\sqrt{y_1^2+y_2^2}$ increases (see Fig.~\ref{fig:flow}(b)). In this way, the flow field obtained by the numerical analysis of the BGK model clearly indicates the presence of a source-sink flow pattern in the far field. This becomes the source-sink condition near the point of discontinuity when rescaled with the spatial variables $x_i$ and the limit $\varepsilon \to 0$ is approached, as discussed in the previous section.

\section{Discussions}
We have considered a slightly rarefied gas confined between two parallel plates whose common temperature distribution has a jump discontinuity along them. In the case of a smooth temperature distribution without the jump discontinuity, the Hilbert expansion and the Knudsen-layer correction yield a practical tool (i.e., the Stokes system) to investigate a thermally-driven flow between the two plates (Sect.~\ref{sec:smooth}). On the other hand, the case of the discontinuous surface temperature cannot be handled solely by the Hilbert solution and the Knudsen-layer correction. Indeed, the term $\phi_{\mathrm{HK}1}$ can grow indefinitely near the point of discontinuity, which disproves the validity of the HK solution there (Sect.~\ref{sec:discontinuou}). 
Given this observation, we have introduced
the Knudsen zone near the point $(x_1,x_2)=(-\frac{\pi}{2},0)$, in which the solution is allowed to undergo an abrupt spatial variation in both $x_1$ and $x_2$ directions.

The Knudsen zone is described by the system \eqref{e:KnZone}, which is a half-space problem for the linearized Boltzmann equation in two space dimensions. In this problem, the constant $\Gamma_z^{(1)}$ occurring in the far-field asymptotic property~\eqref{e:mc_phi_Z0} is essential from the macroscopic view points. Indeed, $\Gamma_z^{(1)}$ is inherited to the source-sink condition \eqref{e:source-sink_2} in the Stokes system and plays a role to induce a non-zero flow velocity $u_{i\mathrm{H}1}$. In this sense, $\Gamma_z^{(1)}$ is of equal importance as the viscosity or the slip/jump coefficients. 

Finally, let us make a brief comment on the global flow structure when $\varepsilon$ is small.
Since the zeroth-order flow velocity $u_{i\mathrm{H}0}$ is identically zero, the overall flow vanishes as $\varepsilon$ tends to zero except in the Knudsen zone. In the Knudsen zone, the nonzero flow of the order $\tauw O(1)$ is induced as seen from Fig.~\ref{fig:flow} and remains. However, the Knudsen zone shrinks to $(x_1,x_2)=(-\frac{\pi}{2},0)$ with the decrease of $\varepsilon$. Therefore, the strong flow of $\tauw O(1)$ is gradually localized near $(x_1,x_2)=(-\frac{\pi}{2},0)$ as $\varepsilon$ becomes smaller. The localized flow affects the global flow at the order $\varepsilon$ through the source-sink condition for $u_{i\mathrm{H}1}$ and induces an overall flow with the magnitude $\tauw O(\varepsilon)$.
In this way, a global flow of the order $\tauw O(\varepsilon)$ is established as a result of the piecewise uniform temperature distribution of the plates. The present analysis successfully provides a clear picture of the flow structure, which is also consistent with the picture inferred in \cite{Aoki-Takata-Aikawa-Golse_PHF97}.





\begin{acknowledgement}
The present work was supported by JSPS KAKENHI Grant No.~17K06146.
\end{acknowledgement}
\section*{Appendix}
\addcontentsline{toc}{section}{Appendix}
In this appendix, we briefly explain the derivation of the condition~\eqref{e:mc_phi_Z0}.
Our stating point is the asymptotic behaviors of the leading order HK solution $\phi_{\mathrm{HK}} = \phi_{\mathrm{HK}0} = \phi_{\mathrm{H}0}$ near $(x_1,x_2)=(-\frac{\pi}{2},0)$, i.e.,
\begin{align}
\phi_{\mathrm{HK}0} & = \frac{2 \tauw}{\pi} \left(|\vec{\zeta}|^2 - \frac{5}{2} \right) \theta + O(r^2),
\quad r \ll 1, \quad \theta = \text{Arctan}\left(\frac{x_2}{x_1 + \frac{\pi}{2}}\right).
\end{align}
This suggests that the leading-order term of $\phi_{\mathrm{Z}}$ is of the form
\begin{align}
\phi_{\mathrm{Z}0} & = \frac{2 \tauw}{\pi} \left(|\vec{\zeta}|^2 - \frac{5}{2} \right) \theta,
\qquad \text{as} \ \ |\vec{y}| \to \infty, \quad y_1 > 0,
\quad \theta = \text{Arctan}\left(\frac{y_2}{y_1}\right).
\label{e:bc_phiZ0_infinity}
\end{align}
Thus, the problem for $\phi_{\mathrm{Z}0}$ consists of \eqref{e:eq_phiZ0}, \eqref{e:bc_phi_Z0}, and \eqref{e:bc_phiZ0_infinity}. We regard this problem as a kind of ``scattering problem'' and seek a solution with the following asymptotic property \cite{taguchi_tsuji_JFM_2020}:
\begin{align}
\phi_{\mathrm{Z}0} & \to \frac{2 \tauw \Gamma_z^{(1)}}{|\vec{y}|} \zeta_r \sin(2\theta) + \frac{2 \tauw}{\pi} \left(|\vec{\zeta}|^2 - \frac{5}{2} \right) \left(\theta
- \frac{c_1^{(0)} \sin \theta}{|\vec{y}|} \right)
\nonumber \\
& \quad
- \frac{2 \tauw}{\pi} \left(\frac{\zeta_\theta}{|\vec{y}|} A(|\vec{\zeta}|) + \frac{1}{y_2} \varphi_1^{(0)}(y_1,\zeta_1,|\overline{\vec{\zeta}}|) \right),
\quad \text{as} \ \ |\vec{y}| \to \infty,
\end{align}
where $\Gamma_z^{(1)}$ is a constant. Note that the terms inversely proportional to $|\vec{y}|$ represent the ``reaction'' to the imposed external condition~\eqref{e:bc_phiZ0_infinity}.

\newcommand{\noop}[1]{}\hyphenation{Post-Script Sprin-ger}

\end{document}